\documentclass[twocolumn]{aastex631}

\usepackage{graphicx}	
\usepackage{amsmath}

\newcommand{\s}{\,{\rm s}}

\newcommand{\pyr}{\,{\rm yr}^{-1}}

\newcommand{\K}{\,{\rm K}}

\newcommand{\mas}{\,{\rm mas}}

\begin{document}

\title{SGR 1935+2154's Quiet Local Environment: Clues for Its Progenitor}

\correspondingauthor{Wenlang He}
\email{hewlang@smail.nju.edu.cn}
\correspondingauthor{Ping Zhou}
\email{pingzhou@nju.edu.cn}
\correspondingauthor{Bingqiu Chen}
\email{bchen@ynu.edu.cn}

\author[0009-0002-4427-6976]{Wenlang He}
\affiliation{School of Astronomy and Space Science, Nanjing University, 163 Xianlin Avenue, Nanjing 210023, People’s Republic of China}
\affiliation{Key Laboratory of Modern Astronomy and Astrophysics, Nanjing University, Ministry of Education, Nanjing 210023, People’s Republic of China}

\author[0000-0002-5683-822X]{Ping Zhou}
\affiliation{School of Astronomy and Space Science, Nanjing University, 163 Xianlin Avenue, Nanjing 210023, People’s Republic of China}
\affiliation{Key Laboratory of Modern Astronomy and Astrophysics, Nanjing University, Ministry of Education, Nanjing 210023, People’s Republic of China}

\author[0000-0003-2472-4903]{Bingqiu Chen}
\affiliation{South-Western Institute for Astronomy Research, Yunnan University, Kunming, Yunnan 650091, People’s Republic of China}

\begin{abstract}

Magnetars are highly magnetized neutron stars (NSs) whose evolution and radiation are governed by the decay and/or reconfiguration of their magnetic fields. The origin of magnetars remains an open question, with proposed progenitor scenarios including core-collapse (CC) of very massive stars ($\ge 25~M_\odot$) or non-very massive stars ($8<M_*<25~M_\odot$), mergers of stellar systems, and accretion-induced collapse (AIC) of white dwarfs (WDs). Investigating the environments of magnetars can offer valuable clues to this issue. In this work, we study the local (a radius of $0.87^\circ$, $\sim 100$ pc at 6.6 kpc) stellar environment of SGR 1935+2154, which is spatially associated with the supernova remnant (SNR) G57.2+0.8, based on astrometry from Gaia DR3 and multi-band photometry from optical to infrared (IR). We discover that the upper limit of the surface density of massive stars around SGR 1935+2154 is only a quarter of that of the solar neighborhood, where the star formation rate is modest in the Galaxy. This quiet environment implies that the magnetar was likely formed by the CC of either a non-very massive star or a binary merger product rather than the CC of a very massive star. Although alternative channels cannot be excluded, their probabilities may be substantially lower. The studies of magnetars associated with SNRs consistently favor non-very massive progenitors, implying that such progenitors may produce a considerable fraction of magnetars. We also backtrack the trajectories of SGR 1935+2154 and its surrounding stars to search for its potential massive companions, yet no such companions are found.

\end{abstract}


\section{Introduction} \label{sec:intro}

Magnetars are a rare and highly magnetized subclass of NSs \citep[see reviews e.g.,][]{2015RPPh...78k6901T, 2017ARA&A..55..261K}, with surface dipole magnetic fields inferred to reach $10^{14}$--$10^{15}$~G. Identified primarily as Soft Gamma Repeaters (SGRs) or Anomalous X-ray Pulsars (AXPs), they are distinguished from ordinary rotation-powered pulsars by their sporadic high-energy bursts and persistent X-ray emission. Their extreme activity is believed to be powered by the decay and/or rearrangement of ultra-strong magnetic fields, rather than by rotational spin-down, accounting for their recurrent bursts, intermediate flares, and rare giant flares with energies up to $10^{46}$~erg \citep{1979Natur.282..587M, 1980ApJ...237L...1C, 1980ApJ...237L...7E, 1999ApJ...515L...9F, 1999Natur.397...41H, 1999AstL...25..635M, 2005Natur.434.1098H, 2005Natur.434.1107P}.

Only about thirty magnetars are currently known in the Galaxy \citep{2014ApJS..212....6O}\footnote{\url{https://www.physics.mcgill.ca/~pulsar/magnetar/main.html}}. Among them, SGR~1935+2154 has emerged as an interesting source. Discovered in 2014 following multiple SGR-like bursts \citep{2014GCN.16520....1S, 2016MNRAS.457.3448I}, it later produced fast radio burst (FRB)~200428—a millisecond-duration radio burst \citep{2020Natur.587...59B, 2020Natur.587...54C} temporally coincident with a hard X-ray burst \citep{2020ApJ...898L..29M,2020ATel13686....1T,2020GCN.27669....1R,2021NatAs...5..378L}. This event provided the first direct evidence linking magnetar activity to FRBs, bridging Galactic magnetars with extragalactic FRB phenomena. Moreover, SGR~1935+2154 is spatially associated with the SNR G57.2+0.8 \citep{2014GCN.16533....1G}, offering a rare opportunity to investigate the surrounding environment and thereby constrain its progenitor properties. Given these characteristics, SGR~1935+2154 represents an exceptional laboratory for testing models of magnetar formation, magnetic field evolution, and high-energy emission mechanisms.

The leading theoretical scenarios for the origin of the magnetar magnetic field invoke either a dynamo mechanism within the proto-neutron star (PNS)—typically requiring extremely rapid rotation (periods of a few milliseconds or less) and vigorous convective motions \citep{1992ApJ...392L...9D}—or a fossil-field hypothesis, where the progenitor possesses a strong magnetic field that is amplified by flux conservation during collapse \citep{2006MNRAS.367.1323F, 2009MNRAS.396..878H}. We distinguish two broad stellar evolutionary pathways to create magnetars: single stars and binary stars. Single stars here are those that do not experience significant interaction with a companion (if exists). In such stars, a fossil field may directly produce a magnetar, or a sufficiently rapidly rotating core may enable dynamo amplification. On the other hand, binary stars here are those that experience interaction or even merge. The spin-up of a star due to binary interaction, e.g., merger, accretion mass transfer, and tidal synchronization, can promote the formation of a magnetar \citep{2016A&AT...29..183P}. In addition, binary NS mergers \citep[e.g.,][]{2013ApJ...771L..26G}, binary WD mergers \citep[e.g.,][]{2006MNRAS.368L...1L}, NS-WD mergers \citep[e.g.,][]{2020ApJ...893....9Z}, and AIC of WDs \citep[e.g.,][]{1991ApJ...367L..19N} are also possible channels for magnetar formation. For recent analyses on magnetar formation channels, we refer the reader to \citet{2025arXiv251106554H}.

To help understand the formation channels of magnetars, many studies have investigated their surrounding environments. One approach is to study the star clusters associated with magnetars. Beyond simple positional coincidence, the association between magnetars and star clusters is supported by independent environmental and kinematic evidence: an IR ring around SGR 1900+14, interpreted as the result of dust destruction by a giant flare from the magnetar, which resides in a cluster of massive stars \citep{2008Natur.453..626W}, and proper motion measurements showing that magnetars such as SGR 1806–20 and SGR 1900+14 are moving away from nearby massive clusters, consistent with an origin within them \citep{2012ApJ...761...76T}. Assuming that the cluster and the magnetar progenitor are of the same age, or alternatively identifying the candidate former companion of a magnetar and analyzing their binary evolution, allows an estimate of the magnetar’s progenitor mass. Early studies on this track support progenitor masses for SGR 1806-20 and CXOU J164710.2-455216 exceeding $\sim 40\rm\,M_{\odot}$ \citep{2005ApJ...622L..49F, 2006ApJ...636L..41M, 2008MNRAS.386L..23B}. However, follow up studies on SGR 1900+14 \citep{2009ApJ...707..844D} and CXOU J164710.2-455216 \citep{2010int..workE..91K, 2020MNRAS.492.2497A} challenge the view that very massive progenitors are the only pathway to magnetar formation. Therefore, the progenitor masses of magnetars associated with star clusters are not necessarily very high. Another approach is to study the SNRs associated with magnetars. Analyses of a few magnetar-associated SNRs indicate normal or low SN explosion energies, suggesting that magnetars are not necessarily born with rapid spin \citep{2006MNRAS.370L..14V,2014MNRAS.444.2910M,2019A&A...629A..51Z}. Studies of Kes~73 (1E~1841$-$045), RCW~103 (1E~161348$-$5055), and N49 (SGR~0526$-$66) suggest that their progenitors were non-very massive, roughly 10--20~$M_\odot$, based on X-ray spectroscopy, elemental abundances, and explosion modeling \citep{2019A&A...629A..51Z, 2023ApJ...950..137N}. A notable exception is PSR~J1119$-$6127, a high-magnetic-field radio pulsar (not a formally recognized magnetar), whose progenitor may have been more massive \citep[$\sim 30~M_\odot$,][]{2012ApJ...754...96K}. Excluding this outlier, SNR-based studies present a remarkably consistent picture: most magnetars associated with SNRs appear to originate from non-very massive progenitors rather than from very massive ones. The association of SGR~1935+2154 with SNR~G57.2+0.8, along with the surrounding stellar environment that remains largely unexplored, thus provides a valuable new case for testing this picture, allowing further investigation into the progenitor properties and evolutionary pathways that give rise to magnetars.

The main goal of this work is to reveal the quiet local stellar environment of SGR 1935+2154, suggesting that it was likely formed by the CC of either a non-very massive star or a binary merger product. In Section \ref{sec:Data}, we describe the data used to quantify the stellar environment of SGR 1935+2154, including astrometry, photometry, and the extinction at SGR 1935+2154. We introduce the stellar atmosphere models used in spectral energy distribution (SED) fitting, and estimate the completeness on each band included in SED fitting. In Section \ref{sec: Methods}, we detail our process for performing SED fitting, from which stellar parameters are derived, and put forward our core idea for identifying massive star candidates around SGR 1935+2154. Section \ref{sec: Results} presents the quiet local environment around SGR~1935+2154 and shows that no massive companion candidates are found. The discussion and conclusion are presented in Sections \ref{sec:Discussion & Conclusion}.

\section{Data} \label{sec:Data}

\subsection{Astrometry}
\label{sec:Astrometry} 

SGR 1935+2154 is located close to the Galactic plane, and its IR counterpart has been identified with coordinates R.A. = $19^{\rm h}34^{\rm m}55^{\rm s}.606$, decl. = $21^{\circ}53^{\prime}47^{\prime \prime}.45$ \citep[$\pm 0.2^{\prime \prime}$,][]{2018ApJ...854..161L}. 
Numerous studies have attempted to constrain the distance to SGR 1935+2154, yielding a wide range of estimates due to differing methodologies and observational constraints. \citet{2016MNRAS.460.2008K} estimate the distance to be $< 10.0\rm\, kpc$ from the burst spectra fitting, \citet{2018ApJ...852...54K} report a distance of $12.5\pm 1.5 \rm\, kpc$ and set a lower limit of $4.5\rm\, kpc$ from the H I structure of the host of SGR 1935+2154 -- SNR G57.2+0.8, and \citet{2020ApJ...905...99Z} derive the distance measurement of $6.6\pm0.7$ of SNR G57.2+0.8 through the local standard of rest (LSR) velocity of molecular clouds that impact the SNR. In our work, we adopt the distance measurement of $6.6\pm0.7$ kpc as a reference. For convenience, we use $d_{\rm \,6.6 kpc}$ to denote the distance normalized to 6.6 kpc. Regarding the proper motion of SGR 1935+2154, we adopt the measurements of SGR 1935+2154 from \citet{2022ApJ...926..121L}, who determined the proper motion of its IR counterpart using observations with the Hubble Space Telescope. Their measurements yield a peculiar proper motion of $\rm{pmRA}=-0.73\pm 0.74\,\mas \pyr, \, \rm{pmDE}=3.03 \pm 1.55 \,\mas \pyr$ with respect to the LSR. 

To study the local stellar environment of SGR 1935+2154, we search for stars within a $0.87^\circ$ radius ($\sim 100\, d_{\rm \,6.6 kpc}\rm\, pc$) centered around the IR counterpart of SGR 1935+2154 in the Gaia DR3 catalog \citep{2023A&A...674A...1G}. We adopt this search radius based on the fact that stars typically remain spatially connected to their siblings within about 100 pc for around 100 Myr even without gravitational binding \citep{2006MNRAS.369L...9B}. Therefore, if the progenitor of SGR 1935+2154 was a very massive star ($ \geq 25\,\rm M_{\odot}$) formed in situ, whose lifetime is less than $\sim 7\, \rm Myr$ \citep{1998A&A...334..505P}, coeval siblings of it would be well within our search radius. Of course, if the progenitor of SGR1935+2154 is a runaway star, the search radius will be much larger than what we adopt here (see Sec. \ref{sec:Discussion & Conclusion} for discussion). A source is kept if either of its distance estimate ranges from \citet{2021AJ....161..147B}, geometric or photogeometric, overlaps with $4.5\rm\, kpc$ to $8.7\rm\, kpc$, where $4.5\rm\, kpc$ and $8.7\rm\, kpc$ are the $3\sigma$ lower and upper distance estimates of SGR 1935+2154, respectively \citep{2020ApJ...905...99Z}. Applying the distance criterion reduces the count from 379,708 to 258,261\footnote{If sources were retained when the 84th percentile of either their geometric or photogeometric distance exceeded $4.5\rm\,kpc$, this selection would yield $258,307$ sources, slightly more than $258,261$.}.
 
\subsection{Photometry}
\label{sec: photometry}
Multi-band photometry is required to perform SED fitting to obtain stellar parameters.

From Gaia DR3 \citep{2023A&A...674A...1G}, we collect $G, BP,RP$-band photometry, and then utilize pre-computed cross matches \citep{2017A&A...607A.105M, 2019A&A...621A.144M}\footnote{\url{https://gea.esac.esa.int/archive/documentation/GDR3/Gaia_archive/chap_datamodel/sec_dm_cross-matches/}} between Gaia DR3 and Pan-STARRS1 (PS1) DR1 \citep[$g,r,i,z,y$-band,][]{2016arXiv161205560C}\footnote{We retrieve photometry from PS1 DR2 based on the fact that the objID of PS1 DR1 is the same as PS1 DR2.}, AllWISE \citep[$W_1$-band,][]{2014yCat.2328....0C}, and 2MASS \citep[$J,H,K$-band,][]{2003yCat.2246....0C}. In addition, we employ the CDS Upload X-Match tool available in TOPCAT \citep{2005ASPC..347...29T} to perform cross-match between Gaia DR3 and UKIDSS \citep[$J_{U},H_{U},K_{U}$-band,][]{2008MNRAS.391..136L} and IGAPS \citep[$i_{I}, H_{\alpha, {I}}, r_{I}$-band,][]{2020A&A...638A..18M} by setting radius to $1^{\prime \prime}$ and pick the closest match.

The photometric uncertainties are the square root of the sum of squared measurement uncertainties and squared systematic errors. The value of the latter is 0.01 mag for Gaia DR3 \citep{2021A&A...649A...3R}, 0.02 mag for PS1 DR2 \citep{2012ApJ...750...99T}, 0.015 mag for AllWISE \citep{2010AJ....140.1868W}, 0.01 mag for 2MASS \citep{2006AJ....131.1163S}, 0.02 mag for UKIDSS \citep{2008MNRAS.391..136L, 2009MNRAS.394..675H}, and 0.02 mag for IGAPS \citep{2020A&A...638A..18M}. 

We discarded unreliable photometry. For Gaia DR3, we keep all the $G$-band data while setting retention criteria for $BP$ and $RP$-band mainly based on \citet{2021A&A...649A...3R}. The retention criteria for $BP$ and $RP$ are presented in Appendix \ref{sec: appendix A}, and the goal of these criteria is to preserve reliable $BP$ and $RP$-band data as much as possible. For PS1 DR2, we remove data whose `qualityFlag' contains 1, 64, or 128, which represent extended objects, suspect objects in the stack, or poor-quality stack objects, respectively \citep{2020ApJS..251....7F}. For AllWISE, we retain data that is unaffected by known artifacts (cc\_flags $=$ 0) and not saturated \citep{2014yCat.2328....0C}. For 2MASS, a quality flag of A, B, or C is required on each band \citep{2003yCat.2246....0C}. For UKIDSS, we impose ppErrbits $< 256$ on each band, and exclude $J_{U} < 13.25$, $H_{U} < 12.75$, and $K_{U} < 12$ photometry data, which risk saturation \citep{2008MNRAS.391..136L}. For IGAPS, photometry is considered reliable if saturation is avoided and the associated class identifies the object as a star or a probable star \citep{2020A&A...638A..18M}.

Finally, to ensure the reliability of the SED fitting, we require photometric measurement from at least 5 bands, including at least one IR band (i.e., $z, y, W_1, J, H, K, J_{U}, H_{U}, K_{U}$). To avoid double-counting a given band in subsequent SED fitting—which would artificially increase its weight—we retain only a single measurement for similar bands. In detail, for the $J$ ($J_{U}$), $H$ ($H_{U}$), and $K$ ($K_{U}$) bands, UKIDSS data should be preferred when available, as it offers deeper sensitivity and higher spatial resolution compared to 2MASS \citep{2006AJ....131.1163S, 2008MNRAS.391..136L}. Similarly, for the $r$ ($r_I$) and $i$ ($i_I$) bands, PS1 DR2 should be preferred compared to IGAPS \citep{2016arXiv161205560C, 2020ApJS..251....7F, 2020A&A...638A..18M}. Hence, we primarily use the bands $G, BP, RP, g, r, i, z, y, J_{U}, H_{U}, K_{U}, W_1,$ and $H_{\alpha, {I}}$. Where $J_{U}, H_{U}$, $K_{U}$, $r$, or $i$ data are absent, we substitute the corresponding $J, H$, $K$, $r_I$, or $i_I$. Finally, there are $240,472$ samples that satisfy the conditions.

\subsection{The Extinction at SGR 1935+2154}
\label{sec: SGR extinction}
Since the extinction $(A_{\rm V})$ toward SGR 1935+2154 is significant, we must account for it when performing SED fitting. \citet{2022ApJ...925L..16Z} reports a value of 5.8 mag using the \citet{2019ApJ...887...93G} interstellar dust map, the distance ${4.4}^{+2.8}_{-{1.3}}$ kpc from \citet{2020ApJ...898L..29M}, and the relations between reddening and extinction from \citet{2013MNRAS.430.2188Y}. On the other hand, \citet{2020ApJ...901L...7D} infer $A_{\rm V}=7.2\pm 0.9$ mag from the neutral hydrogen column density along the line of sight from XMM-Newton spectra fitting \citep{2016MNRAS.457.3448I} and using the relation between optical extinction and hydrogen column density from \citet{2009MNRAS.400.2050G}. Taken together, the extinction at SGR 1935+2154 falls roughly within the range of 5.8 to 8.1 $(7.2+0.9)$ mag. We note that neither of these represents a direct measurement of $A_{\rm V}$.

\subsection{Stellar Atmosphere Models}
\label{sec: Model Atmospheres} 
Stellar atmosphere models are essential for SED fitting. Given that we are concerned with massive stars, we have chosen the TLUSTY model \citep{2003ApJS..146..417L, 2007ApJS..169...83L}. We download the model from SVO service\footnote{\url{http://svo2.cab.inta-csic.es/theory/newov2/index.php?models=tlusty_mergedbin}}. The ranges of stellar parameter space are effective temperature $T_{\mathrm{eff}} = 15,000 -55,000$ K, surface gravity $\log g = 1.75-4.75$, and metallicity
$Z = 0-2$ $Z_{\odot}$. Given that metallicity has little impact on our SED fitting, we fix $Z = 1$ $Z_{\odot}$. 

To assess completeness (Sec. \ref{sec: Completeness}), we also make use of the Kurucz model \citep{2003IAUS..210P.A20C} downloaded from SVO service\footnote{\url{http://svo2.cab.inta-csic.es/theory/newov2/index.php?models=Kurucz2003}}, since the stellar parameter space is much broader than that of TLUSTY model. The ranges of stellar parameter space are $T_{\mathrm{eff}} = 3,500 -50,000$ K, $\log g = 0-5$, and 
$Z = 10^{-2.5}-10^{0.5}$ $Z_{\odot}$. Again, we fix $Z = 1$ $Z_{\odot}$. 

\subsection{Completeness}
\label{sec: Completeness} 

To assess the observational completeness of our sample, we estimate the minimum detectable initial mass of a star towards SGR 1935+2154. Three ingredients are needed: the completeness limits on each band, the expected absolute magnitude ($M_{\text{abs}}$) of a star of a given initial mass at a certain age, and the extinction on each band.

The completeness limits are $G=21.0, BP=20.3, RP=20.3$ \citep[Gaia DR3,][]{2021A&A...649A...3R}, $g=23.3, r=23.2, i=23.1, z=22.3, y=21.4$ \citep[PS1,][]{2016arXiv161205560C}, $W_1=17.1$ (AllWISE\footnote{\url{https://wise2.ipac.caltech.edu/docs/release/allwise/expsup/sec2_4a.html}}), $J=15.8, H=15.1, K=14.3$ \citep[2MASS,][]{2006AJ....131.1163S}, $J_U=18.5, H_U=18.0, K_U=17.5$ \citep[UKIDSS, Figure 1 in][]{2008MNRAS.391..136L}, $i_{I} = 20.4, H_{\alpha, {I}} = 20.5, r_{I} = 21.5$ \citep[IGAPS,][]{2020A&A...638A..18M}.

The $M_{\text{abs}}$ of a star of a given initial mass is estimated using MIST \citep[MESA Isochrones \& Stellar Tracks,][]{2011ApJS..192....3P,2016ApJS..222....8D, 2016ApJ...823..102C}. We download evolutionary tracks for multiple synthetic photometry using ``Web Interpolator'' from MIST  Home\footnote{\url{https://waps.cfa.harvard.edu/MIST/interp_tracks.html}}. Specifically, we set Initial v/v\_crit = 0.4, [Fe/H] = 0, and the initial mass values matching those in the ``Packaged Model Grids''\footnote{\url{https://waps.cfa.harvard.edu/MIST/model_grids.html}}. A star's $M_{\text{abs}}$ keeps changing as it evolves. To have a reference, we use the zero age main sequence (ZAMS), which generally marks the dimmest point in a star’s life, apart from its final stages.

The third part concerns the extinction in each band, including the factors used to convert $A_{\rm V}$ into other bands. Since these conversion factors vary with stellar spectra, they are particularly sensitive for some broad bands such as the $G$ band. We thus adopt the Kurucz model to calculate conversion factors for various combinations of $(T_{\mathrm{eff}}, \log g, A_{\rm V})$, where $A_{\rm V}$ is from 0 to 10 mag with increments of 0.5 mag. Using the Python package extinction \citep{2016zndo....804967B}\footnote{\url{https://extinction.readthedocs.io/en/latest/}}, we employ the \citet{1999PASP..111...63F} dust extinction function with ratio of total to selective extinction $R_V=3.1$ and apply it on every wavelength point in stellar atmosphere models. Flux comparisons before and after considering $A_{\rm V}$ provide extinctions in each band for different $(T_{\mathrm{eff}}, \log g, A_{\rm V})$, yielding the corresponding conversion factors. Finally, we construct an interpolator utilizing RBFInterpolator in SciPy \citep{2020NatMe..17..261V} to estimate conversion factors for any combinations of $(T_{\mathrm{eff}}, \log g, A_{\rm V})$. All ranges of input features are scaled to [0, 1], and kernel=`linear' \& neighbors=30 are set for RBFInterpolator.

\begin{figure*}. 
	\centering
		\centering
		\includegraphics[width=1\textwidth]{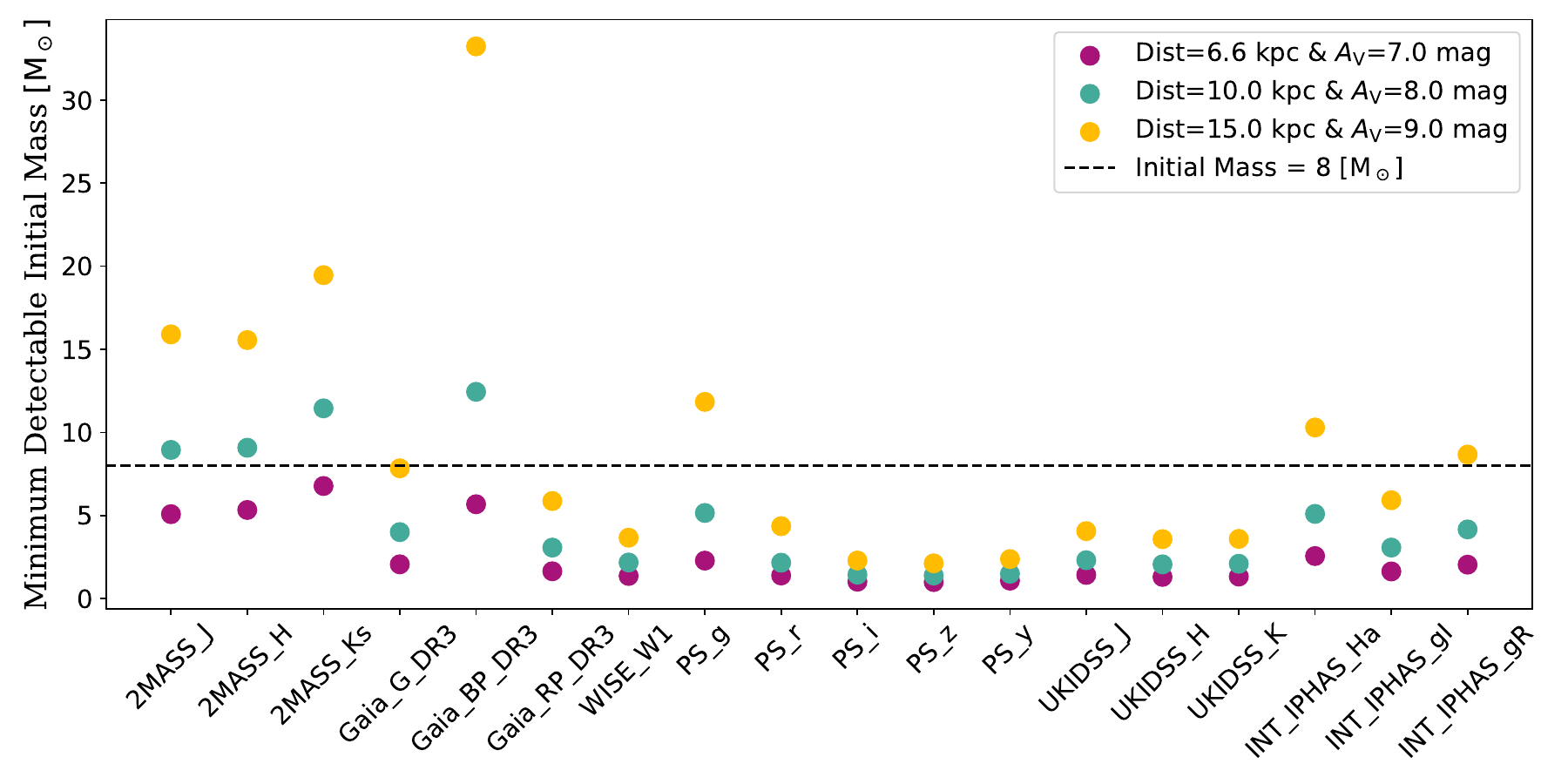}
		\caption{Minimum detectable initial mass towards SGR 1935+2154 in each band. We consider three different combinations of distance and extinction: $(6.6~\mathrm{kpc},~7~\mathrm{mag})$, $(10~\mathrm{kpc},~8~\mathrm{mag})$, and $(15~\mathrm{kpc},~9~\mathrm{mag})$, where $(6.6~\mathrm{kpc},~7~\mathrm{mag})$ represents a plausible location of SGR 1935+2154 (see Sec. \ref{sec:Astrometry} and Sec. \ref{sec: SGR extinction}). Even at $(15~\mathrm{kpc},~9~\mathrm{mag})$, massive stars would in principle still be included in our sample of massive star candidates (see Sec. \ref{sec: photometry}).}
        \label{fig:Minilimit_vs_Bands}
\end{figure*}

Integrating the three ingredients listed above, Figure \ref{fig:Minilimit_vs_Bands} displays the minimum detectable initial mass towards SGR 1935+2154 in each band. In principle, if a star is a massive star located near the SGR 1935+2154, it should be included in our SED-fitting sample (see Sec. \ref{sec: photometry}).

\section{Methods}
\label{sec: Methods}

\subsection{SED Fitting}
\label{sec: SED Fitting} 
In this section, we describe how stellar parameters such as $T_{\rm eff}$ and $A_{\rm V}$ are derived through SED fitting, which we use to identify massive star candidates around SGR 1935+2154.

To derive stellar parameters, we compare the TLUSTY model with multi-band photometry. In doing so, we pre-compute the flux per unit wavelength $(F_{\lambda})$ predicted by TLUSTY model, expressed in units of erg s$^{-1}$ cm$^{-2}$ Å$^{-1}$, in each band for various $(T_{\mathrm{eff}}, \log g, A_{\rm V})$ combinations, where $A_{\rm V}$ spans 0 -- 10 mag in 0.02 mag increments, a wider interval chosen to obtain a more complete sample of massive star candidates (See Sec. \ref{sec: SGR extinction} for the extinction at SGR 1935+2154). In detail, we utilize the Python package pyphot \citep{zenodopyphot}\footnote{\url{https://mfouesneau.github.io/pyphot/index.html}} to obtain the $F_{\lambda}$ in different bands. When the provided filter library\footnote{\url{https://mfouesneau.github.io/pyphot/libcontent.html}} in pyphot includes the relevant band, we use the transmission provided by pyphot. If a band is not available, we download the transmission from the SVO Filter Profile Service \citep{2020sea..confE.182R}\footnote{\url{http://svo2.cab.inta-csic.es/theory/fps/index.php?mode=browse}}. For the treatment of extinction in each band and the construction of the interpolator of $F_{\lambda}$ for arbitrary $(T_{\mathrm{eff}}, \log g, A_{\rm V})$ combinations, we perform an operation similar to that in the penultimate paragraph of Sec. \ref{sec: Completeness}, except that the conversion factor is replaced with $F_{\lambda}$. Finally, a scaling factor is required to scale the atmosphere model to the observed photometry, taking into account the distance $(D)$ and the radius $(R)$ of a star, which we denote as $\alpha (\equiv \frac{R^2}{D^2})$. The range of $\alpha$ is estimated from the ratio of the observation to the model, taking the maximum and minimum of these ratios. Because this range spans orders of magnitude, we work with $\log_{10}\alpha$ rather than $\alpha$.

\begin{figure*}[htbp]
    \centering
    \includegraphics[width=0.3\textwidth]{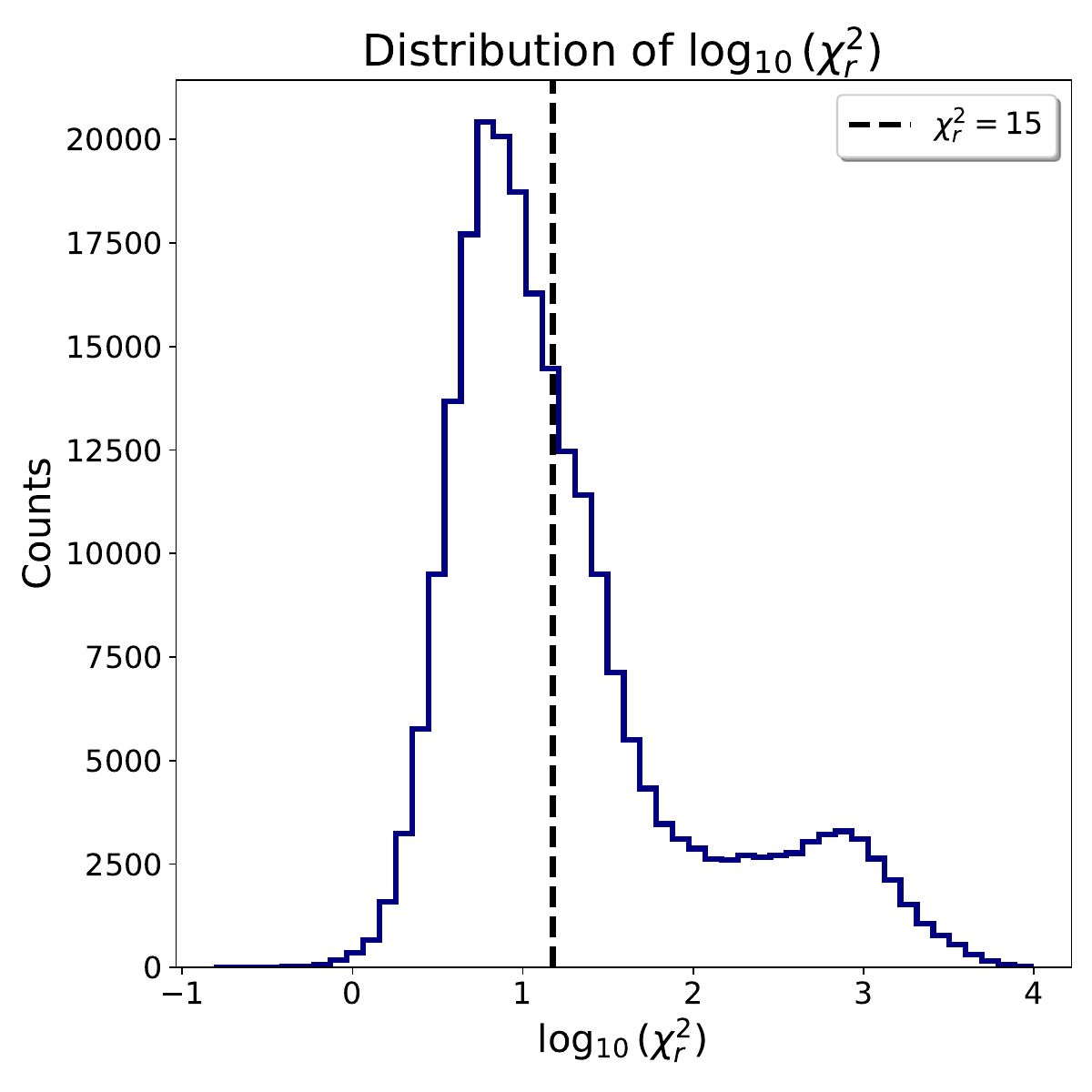}
    \includegraphics[width=0.3\textwidth]{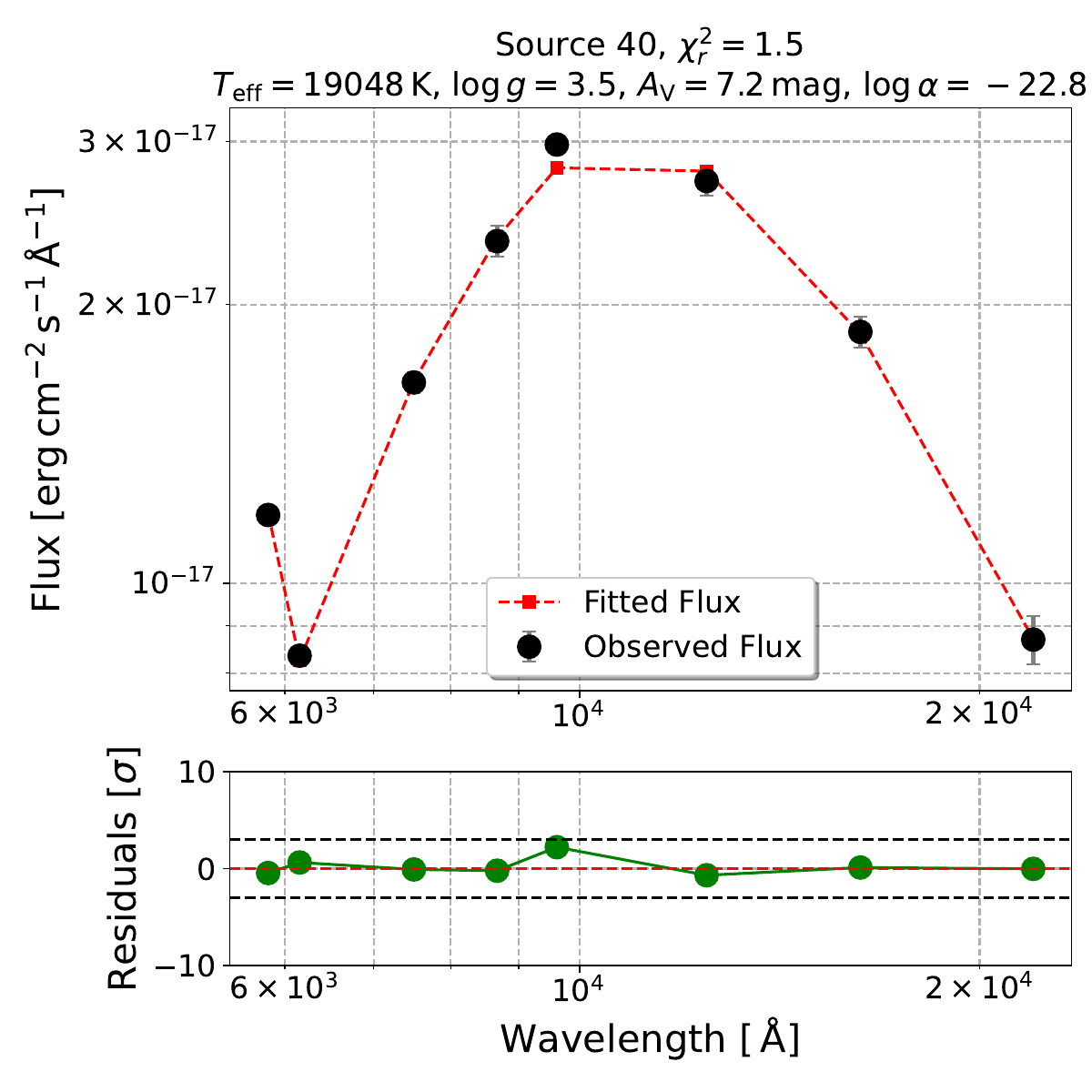}
    \includegraphics[width=0.3\textwidth]{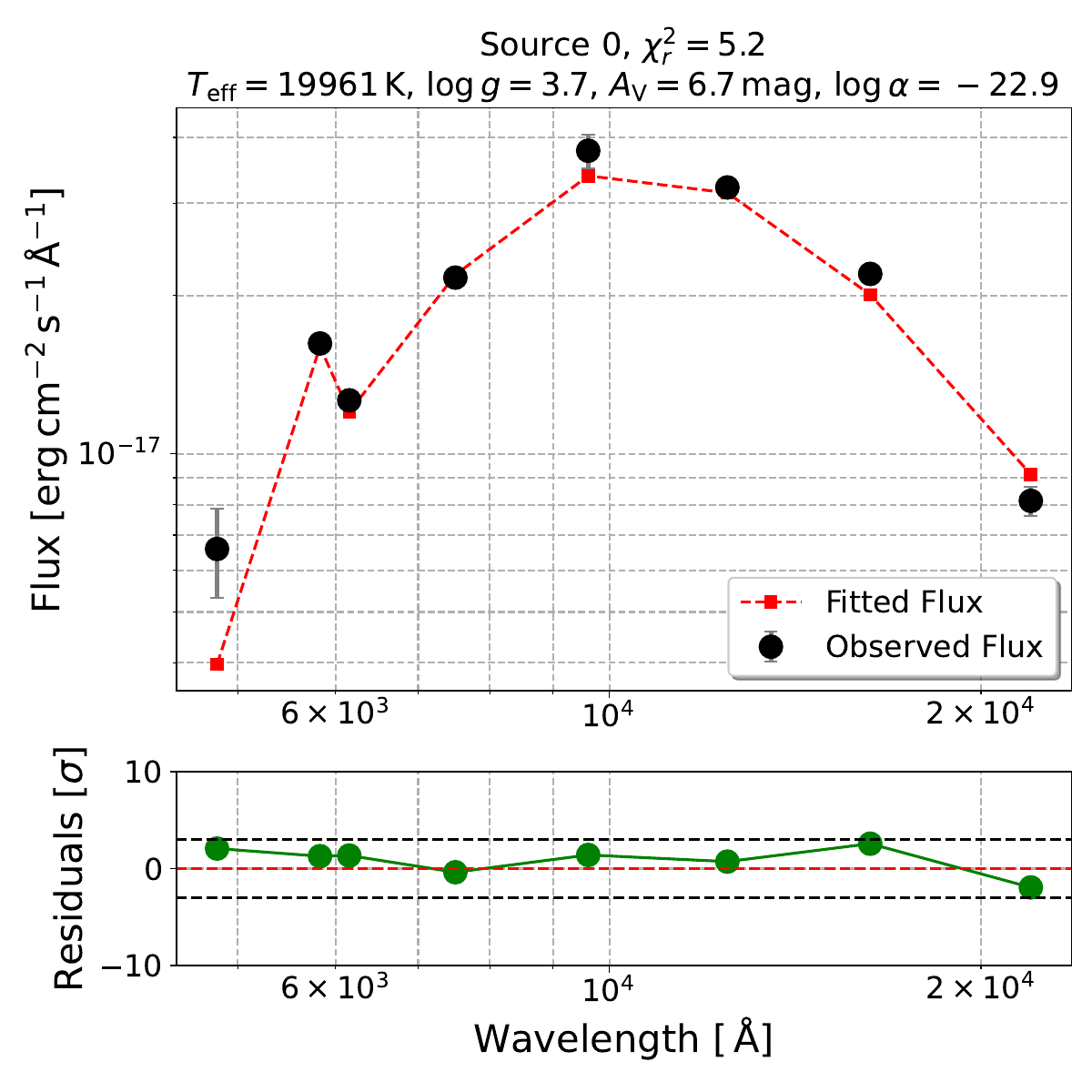}\\[1ex]

    \includegraphics[width=0.3\textwidth]{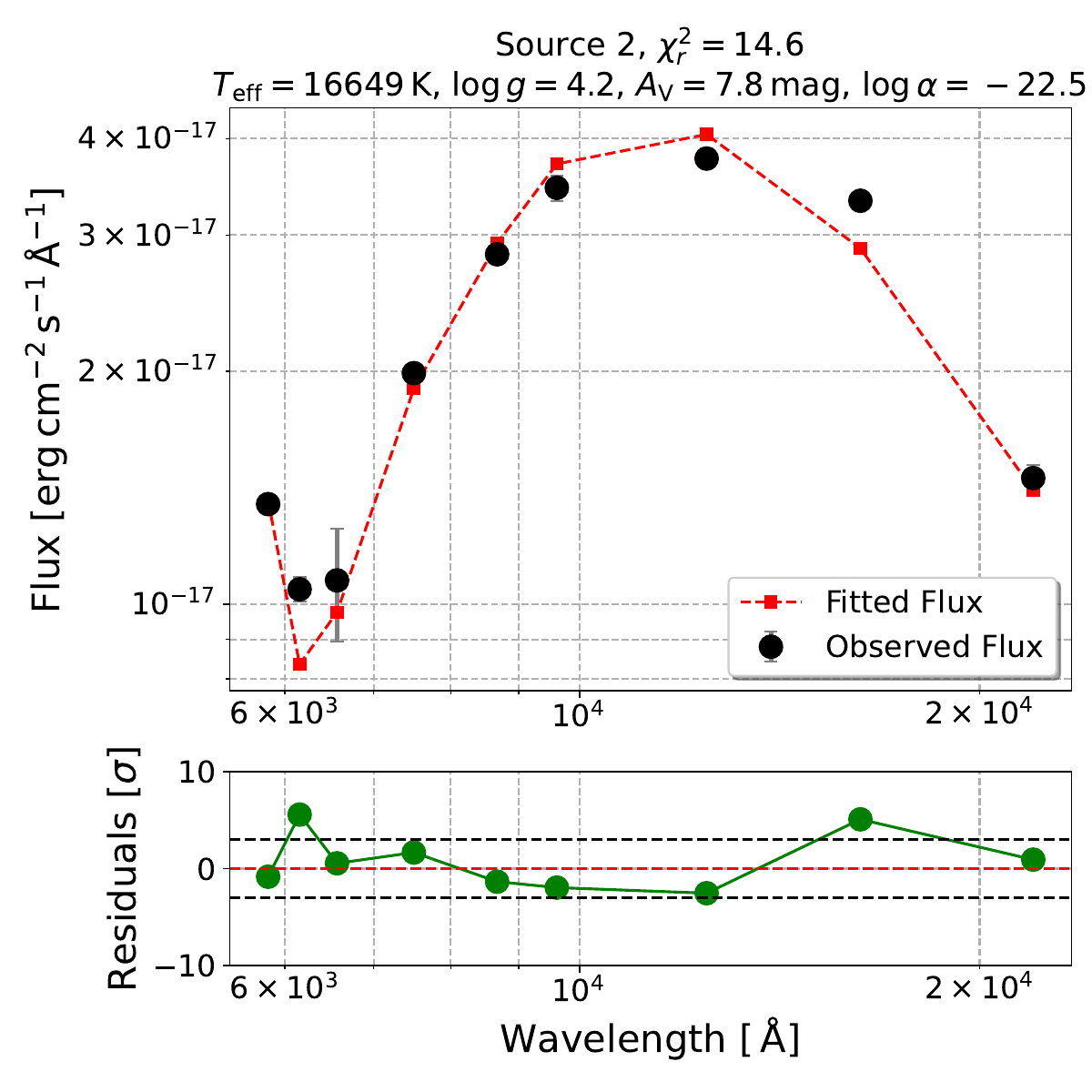}
    \includegraphics[width=0.3\textwidth]{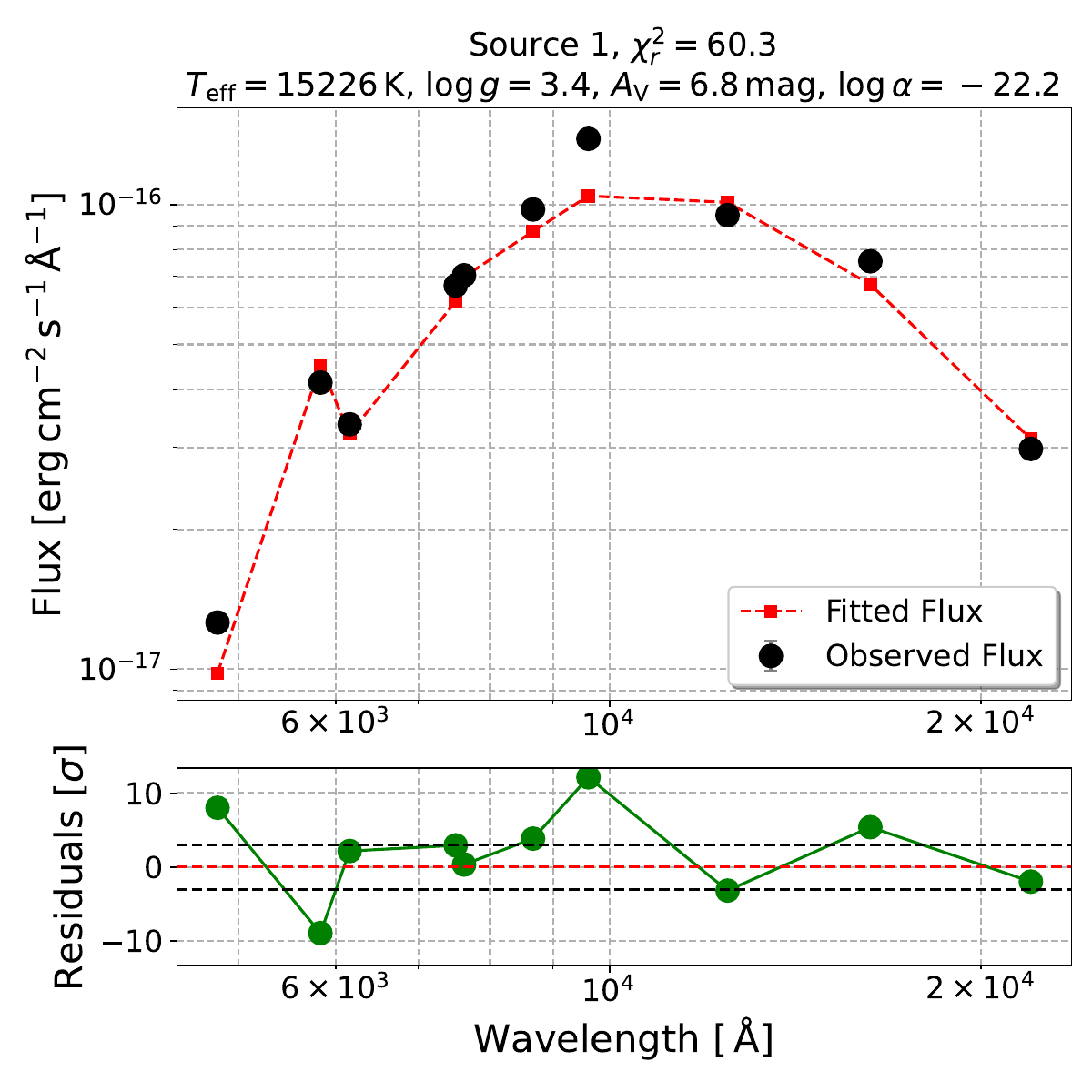}
    \includegraphics[width=0.3\textwidth]{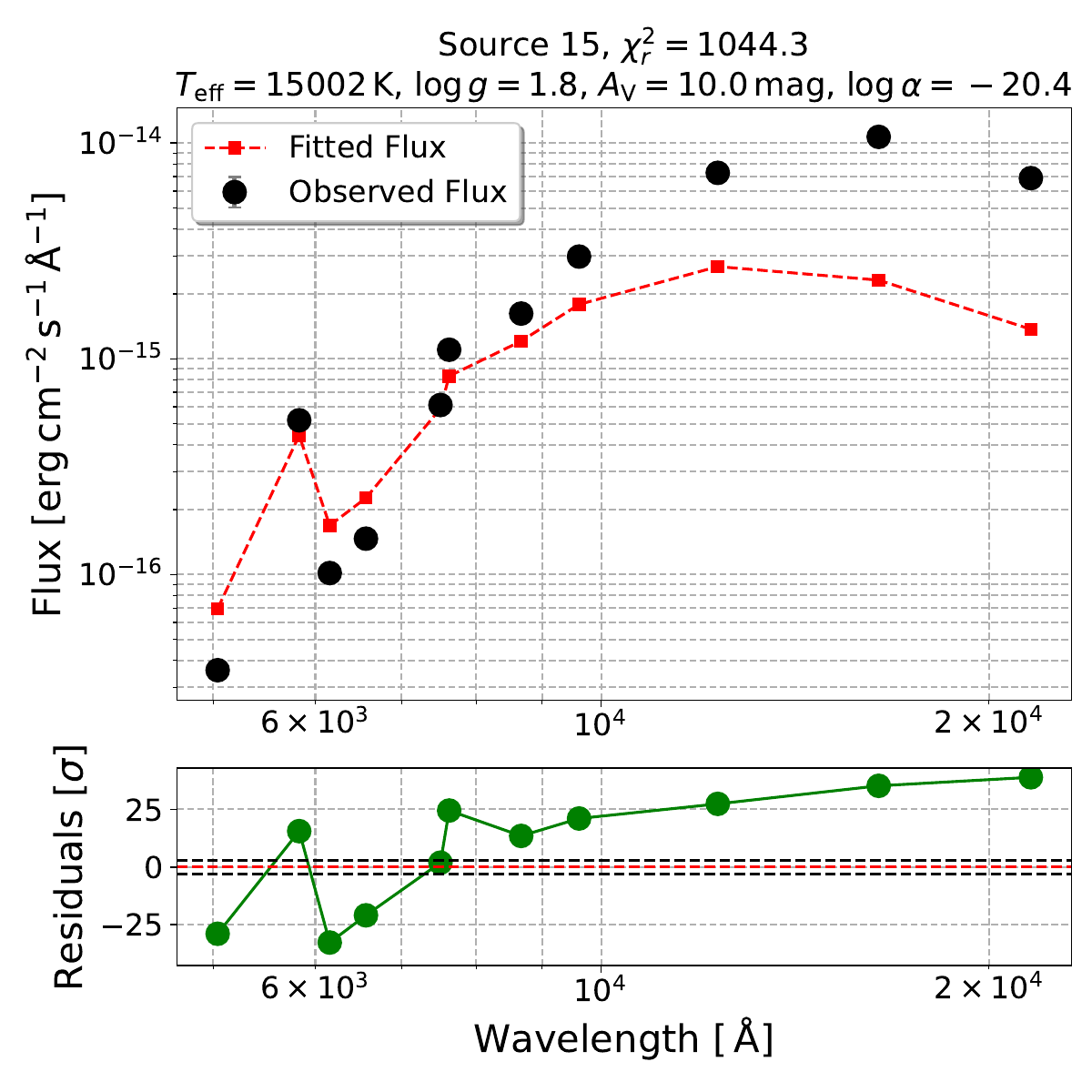}

    \caption{The distribution of reduced chi-square $\chi^2_r$ for the median fitted results (top left) along with the fitted results for different values of $\chi^2_r$ (others). \textit{Top left:} Sources are retained based on the cut $\chi^2_r \leq 15$, as indicated by the black dashed line. A considerable number of sources show large $\chi^2_r$ values since we use TLUSTY models (a hot star model) and $A_{\rm V}$ is limited to $0-10$ mag in order to identify massive star candidates around SGR 1935+2154, so cooler or more highly extincted stars cannot be fitted. \textit{Others:} The fitted results are sorted in ascending order of $\chi^2_r$, along with the corresponding residuals (defined as [observed - fitted] / error) for each band. The two black dashed lines in the residual plots designate the $+3\sigma$ and $-3\sigma$ levels.}
    \label{fig: chisquare}
\end{figure*}

We perform SED fitting through the Markov Chain Monte Carlo (MCMC) method of \citet{2010CAMCS...5...65G}, as implemented in the Python package emcee \citep{2013PASP..125..306F}, for $240,472$ samples that satisfy the conditions described in Sec. \ref{sec: photometry}. The prior is uniformly distributed over the parameter space, and the log-likelihood function is given by
\begin{equation}
\mathcal{L}(\theta) = \sum_{i=1}^{N} \left[ -\ln\left(\sqrt{2\pi} \cdot \sigma_i\right) - \frac{\left(O_i - M_i(\theta)\right)^2}{2\sigma_i^2} \right]
\end{equation}\\
where $\theta$ contains $T_{\mathrm{eff}}$, $\log g$, $A_{\rm V}$, and $\alpha$, $N$ is the total number of valid observed bands, $O_i$ is the $i$-th valid observed $F_{\lambda}$, $M_i(\theta)$ is the $i$-th predicted $F_{\lambda}$, and $\sigma_i$ is the uncertainty of the $i$-th observed $F_{\lambda}$, which takes measurement uncertainty and systematic error into account.
The MCMC chains are initialized with 500 walkers. We used a combination of KDEMove (60\%) and StretchMove (40\%) to balance parameter space exploration and convergence, and set the maximum number of steps to $10,000$. In practice, we find that this number of steps is already enough to judge whether the model can reasonably fit the observation. If chains for each parameter exceed 50 times the autocorrelation time $(\tau)$, we consider the chains to be sufficiently converged, and set a burn-in period of $5\tau$ and thinning by retaining every $0.5\tau$ steps. When convergence is not achieved, we consider that the model cannot reasonably fit the observation and the fitted result is unreliable, and set a burn-in period of $5,000$ steps and thinning by retaining every $20$ steps. To assess the quality of the fit, we compute the reduced chi-square for the median fitted results
\begin{equation}
\chi^2_r = \frac{1}{N - N_p}\sum_{i=1}^{N}\frac{\left(O_i - M_i(\theta)\right)^2}{\sigma_i^2}
\end{equation}\\
where $N_p$ denotes the number of parameters (4: $T_{\mathrm{eff}}$, $\log g$, $A_{\rm V}$, and $\alpha$). Figure \ref{fig: chisquare} displays the distribution of $\chi^2_r$ along with the fitted results for different values of $\chi^2_r$. A considerable number of sources show large $\chi^2_r$ values, as anticipated. This arises from our choice of TLUSTY models—optimized for hot stars—and restrict $A_{\rm V}$ to $0-10$ mag since our focus is on massive stars around SGR 1935+2154, so many cooler or more highly extincted stars are expected to have large $\chi^2_r$ values. Sources with $\chi^2_r \leq 15$ (see the bottom of the left column in Fig. \ref{fig: chisquare} for the fit at $\chi^2_r \sim 15$) are selected for subsequent analysis, resulting in $137,903$ sources. 

\subsection{Searching for Massive Star Candidates}
\label{sec: Massive star candidates} 
The stellar environment around SGR 1935+2154 is crucial for revealing recent in situ star formation activity. In this subsection, we introduce our idea to identify massive star candidates in the absence of reliable distances.

Since distance estimates at large distance ($\gtrsim 3\,\rm kpc$, as in our case) are generally unreliable if based only on Gaia's information, estimate $M_{\text{abs}}$ using $M_{\text{abs}} = m_{\text{app}} - 5 \log_{10}(D) - A + 5$, where $m_{\text{app}}$ is the apparent magnitude, $A$ is the extinction, and $D$ is the distance from \citet{2021AJ....161..147B}, is risky. Instead, we estimate the distance each source would need to have in order to be considered massive. This is our core idea. An alternative method is to estimate stellar mass by interpolating the MIST grids with the SED-derived $T_{\mathrm{eff}}$ and $\log g$. Yet optical to IR photometry alone does not constrain $T_{\mathrm{eff}}$ and $\log g$ well, resulting in a wide mass interval. Therefore, we do not adopt this method.

The $M_{\text{abs}}$ of the ZAMS of a $8\,\rm M_{\odot}$ star serves as a benchmark to determine how far away our sample would have to be if they were massive stars $(\geq 8\,\rm M_{\odot})$. We estimate $M_{\text{abs}}$ following the procedure described in the third paragraph of Sec. \ref{sec: Completeness}. Given that over 99.4\% of $137,903$ sources possess $G$-band measurements with high precision, and that Fig. \ref{fig:Minilimit_vs_Bands} shows massive stars near SGR 1935+2154 should be visible in the $G$ band, we decide to employ $G$-band photometry for identifying massive star candidates. 

The $G$-band extinction $A_G$ is required, which is derived from the SED-derived $T_{\mathrm{eff}}$, $\log g$, and $A_{\rm V}$. In order to construct an interpolator to estimate $A_G$ for any combinations of $(T_{\mathrm{eff}}, \log g, A_{\rm V})$, we perform an operation similar to that in the penultimate paragraph of Sec. \ref{sec: Completeness}, except that we use the TLUSTY model and set the increment of $A_{\rm V}$ to 0.02 mag. We find that $A_{\rm V}$ is the dominant factor in determining $A_G/A_{\rm V}$ in the TLUSTY model (a hot star model), whereas variations in $T_{\mathrm{eff}}$ and $\log g$ have little impact on this ratio. In our SED fitting results, the typical uncertainty of $A_{\rm V}$ is 0.1 mag, with 99\% of the sources having uncertainties within 0.3 mag.
This means that $A_G$ could be reliably estimated.

With the $M_{\text{abs}}$ on $G$ band of the ZAMS of a $8\,\rm M_{\odot}$ star, the $m_{\text{app}}$ from $G$-band observation, and the $A_G$ available, we can estimate how far away a source must be if it is a massive star using the formula $  \log_{10}(D) = 0.2\times (m_{\text{app}}-M_{\text{abs}} - A_G + 5)$.

\section{Results}
\label{sec: Results}

\subsection{No Concentration of Massive Stars in the Vicinity of the SGR 1935+2154}
\label{sec: No Concentration of Massive Stars in the Vicinity of the SGR 1935+2154}
\begin{figure*}
	\centering
		\centering
		\includegraphics[width=0.5\textwidth]{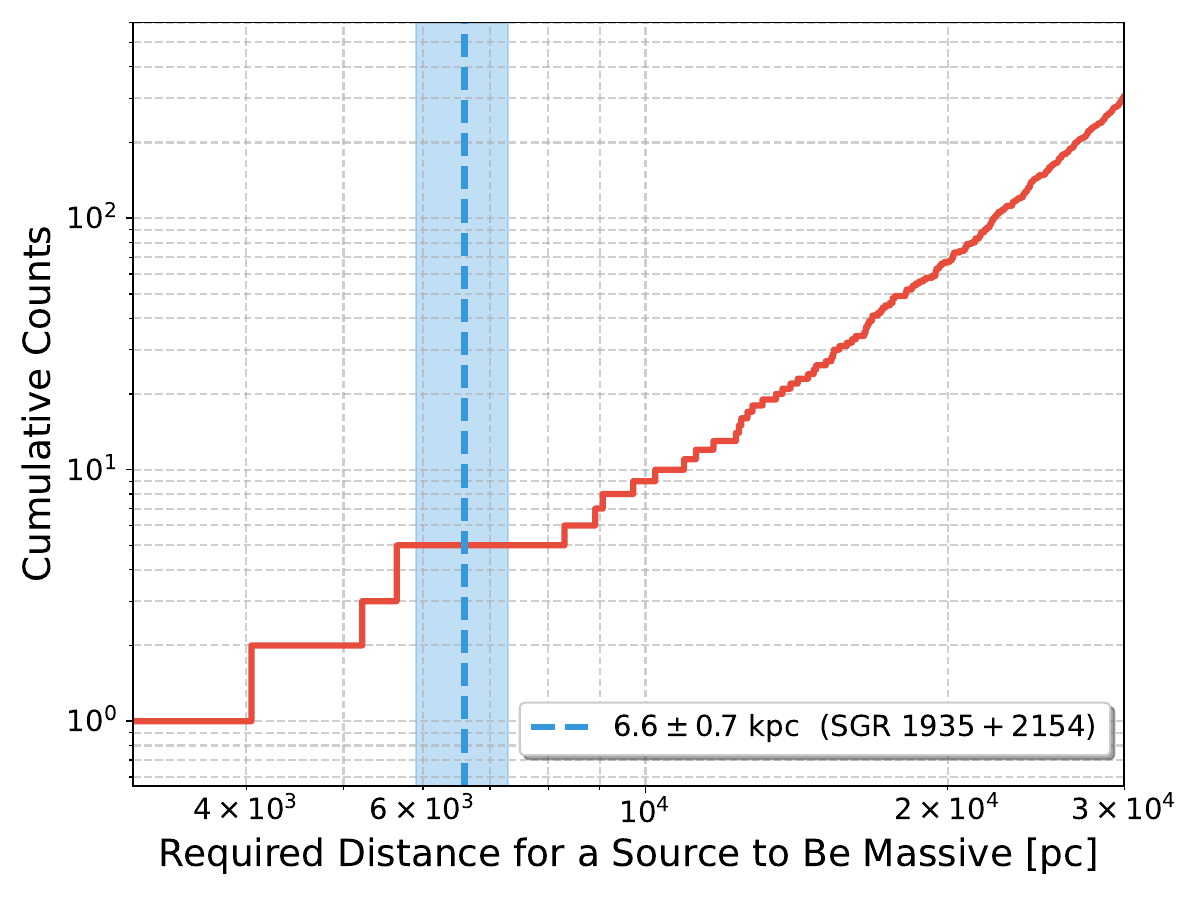}
		\caption{Cumulative counts of the required distances for the sources to be considered as massive star candidates towards SGR 1935+2154. The blue shaded region denotes the reference distance range adopted for SGR 1935+2154 \citep{2020ApJ...905...99Z}. A considerable fraction of the sources would need to be placed at distances far beyond that of SGR 1935+2154, even outside the Galactic disk (we truncate the plot at 30 kpc), to be considered massive star candidates. These sources are therefore very likely not massive stars in the vicinity of the SGR 1935+2154, so they are not the focus of our study.}
		\label{fig:distance_highmass_cdf}
\end{figure*}

Figure \ref{fig:distance_highmass_cdf} is an illustration of the massive star candidates towards SGR 1935+2154. It shows the cumulative counts of the required distances for the sources to be considered as massive star candidates. A maximum of 5 massive star candidates are identified within the adopted distance of 6.6 kpc for SGR 1935+2154, increasing to 9 within 10 kpc and 26 within 15 kpc.

\begin{figure*} 
	\centering
		\centering
		\includegraphics[width=\textwidth]{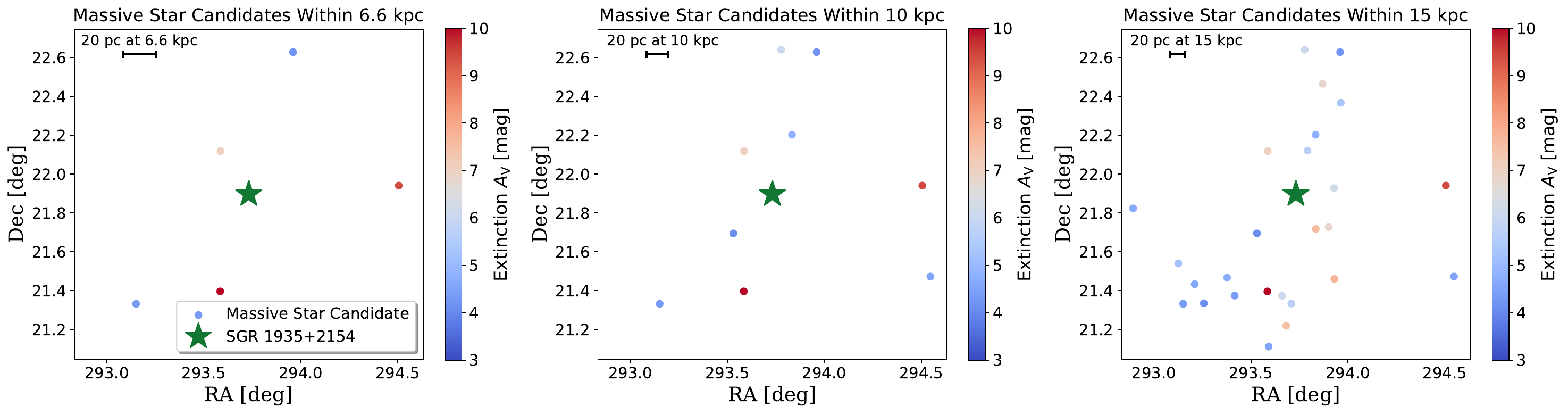}
		\caption{Spatial distribution of massive star candidates around SGR 1935+2154 within $6.6$ kpc (left), $10$ kpc (middle), and $15$ kpc (right). The green star marks the position of SGR 1935+2154, while the circles indicate the massive star candidates, color-coded by their $A_{\rm V}$ values.}
    \label{fig:massive_star_candidates_extinction_distribution}
\end{figure*}

Massive star candidates within $6.6$ kpc (left), $10$ kpc (middle), and $15$ kpc (right) towards SGR 1935+2154 are shown in Figure \ref{fig:massive_star_candidates_extinction_distribution}, with colors representing their $A_{\rm V}$ values. These massive star candidates are sparsely distributed in both space and extinction. The only exception is a clump in the lower-left corner of the right panel, where sources are relatively concentrated in the space–extinction plane. However, since their $A_{\rm V}$ values are small but the required distances are large $(\sim 15 \, \rm kpc)$, these sources are more likely to be foreground non-massive stars rather than genuine massive star candidates. Even when ignoring the extinction variation and considering all candidates together, their surface densities remain low, only about $2 \times 10^{-4}$, $1 \times 10^{-4}$, and $2\times 10^{-4}\,\rm pc^{-2}$ at $6.6$, $10$, and $15$ kpc, respectively.

For comparison, we select the solar neighborhood, whose star formation rate lies at a moderately low level within the Galaxy \citep{2023A&A...678A..95S}. We search the Alma catalogue \citep{2025MNRAS.543...63P} for massive stars (with a ``GLS'' prefix) within 100, 152, and 228 pc (the linear sizes corresponding to our search radius $0.87^\circ$ at distances of 6.6, 10, and 15 kpc
) of the Sun and find 13, 48, and 101 sources, respectively. Their surface densities correspond about $4 \times 10^{-4}$, $7\times 10^{-4}$, and $6\times 10^{-4}\,\rm pc^{-2}$, respectively, which are overall four times higher than the upper limit of that of SGR 1935+2154. We also search the Alma catalogue for massive stars within $0.87^{\circ}$ of SGR 1935+2154 and find only one source locates at a distance of about 1.2 kpc, highly likely reflecting the catalogue's incompleteness. We expect that future updates of Galactic massive star catalogue may allow a more direct comparison with our results.

\subsection{Absence of Massive Companions around the SGR 1935+2154}
\label{sec: Absence of Massive Companions around the SGR 1935+2154}
Massive stars are generally born in binary or even higher-order multiple systems \citep[e.g.,][]{2012Sci...337..444S, 2017ApJS..230...15M}. 
Observational studies have shown that the multiplicity frequency increases with the ZAMS mass of the primary star \citep[][and references therein]{2017ApJS..230...15M}, 
implying that nearly all O-type stars are members of multiple systems. Combining the distribution of the mass ratio of companion to primary and the multiplicity frequency among different orbital periods from Table 13 in \citet{2017ApJS..230...15M}, one can in principle estimate the number of massive companions an O-type star has. If the progenitor of SGR 1935+2154 was a very massive star ($\ge 25~M_\odot$), it may have been part of a binary or multiple system. If the progenitor was the secondary component, the initially more massive primary would have already exploded. Conversely, if the progenitor was the primary, some of its companions could have been massive stars and might still survive provided they did not merge with the progenitor.

To search for such potential companions, we backtrack the trajectories of SGR 1935+2154 and its surrounding stars. We collect proper motions of the sample selected in Sec. \ref{sec:Astrometry} from Gaia DR3 and of SGR 1935+2154 from \citet{2022ApJ...926..121L}. The backtracked time interval is from $10,000$ to $100,000$ yr, covering the age range of SGR 1935+2154 \citep{2018ApJ...852...54K, 2020ApJ...905...99Z}, and the timestep of backtracked time is determined by dividing the precision of angular distance, which is set to better than $3^{\prime \prime}$ ($\sim 0.1\,d_{\rm \,6.6 kpc}\,\rm pc$), by the relative proper motion of the sample with respect to SGR 1935+2154. To quantify uncertainties in the proper motion space, we randomly sample 500 values for each proper motion component of SGR 1935+2154 and its surrounding samples with Gaussian distribution.

\begin{figure*} 
	\centering
		\centering
		\includegraphics[width=0.5\textwidth]{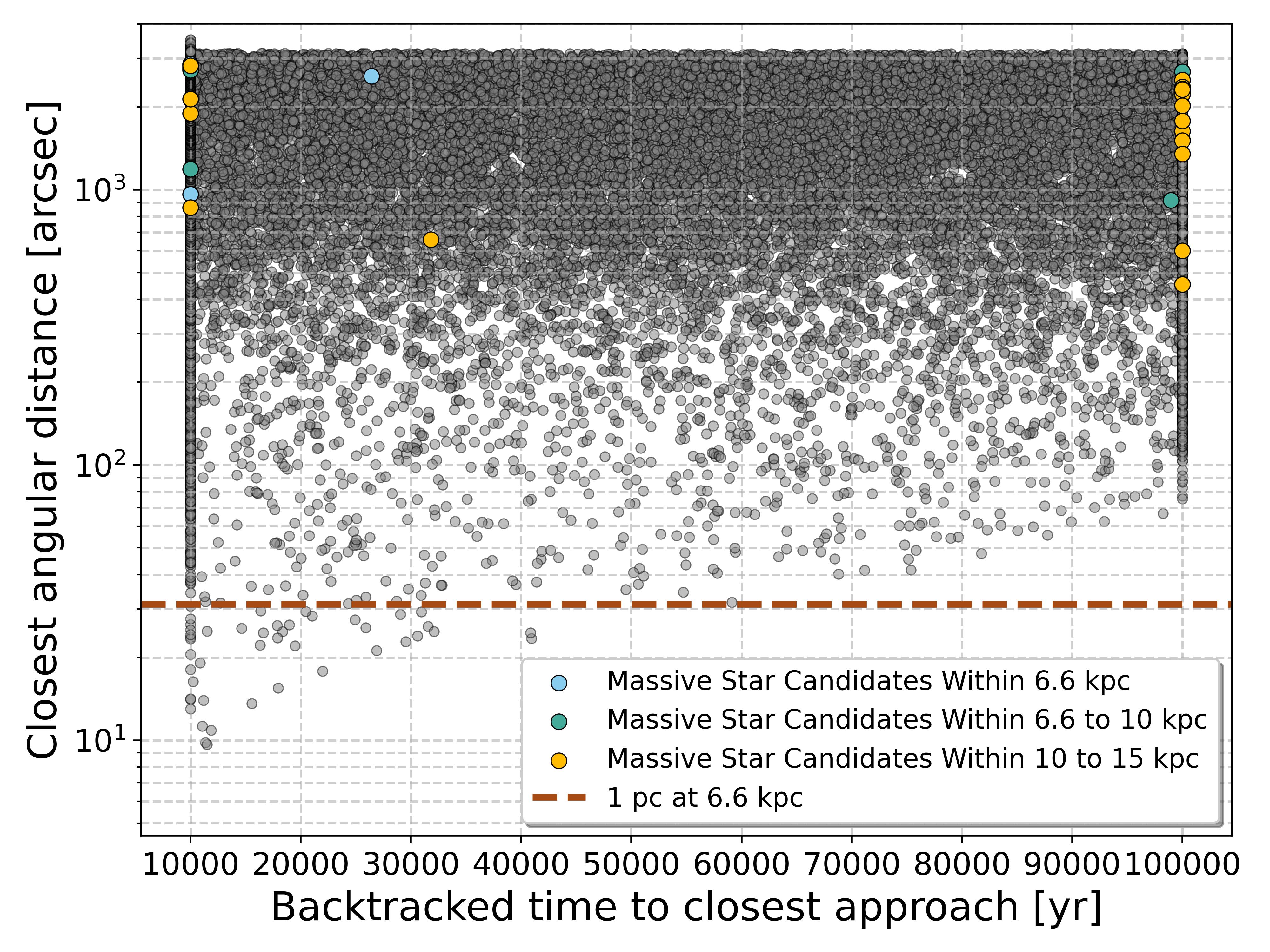}
		\caption{The median results of the backtracked time to closet approach of the samples with respect to SGR 1935+2154 versus the closest angular distance. The gray points in background are all the sample selected in Sect. \ref{sec:Astrometry}. Massive star candidates are marked in blue ($\leq 6.6$ kpc), green (6.6--10 kpc), and orange (10--15 kpc), with the dashed line showing the angular separation subtended by 1 pc at 6.6 kpc.}
        \label{fig:closest_angulardistance_vs_Backtrackedtime}
\end{figure*}

Figure \ref{fig:closest_angulardistance_vs_Backtrackedtime} shows the median results of the backtracked time to closet approach of the samples with respect to SGR 1935+2154 versus the closest angular distance. No massive star candidates were close to SGR 1935+2154 at the time of its birth. For massive star candidates, in the median results, the nearest one lies at a projected separation of about $450^{\prime \prime}$ ($\sim 14\,d_{\rm \,6.6 kpc}\,\rm pc$), while in all sampling results, the nearest separation is about $123^{\prime \prime}$ ($\sim 4\,d_{\rm \,6.6 kpc}\,\rm pc$). Our result is consistent with previous studies \citep{2022MNRAS.513.3550C, 2024MNRAS.531.2379S}, in that no bound or unbound companions of SGR 1935+2154 were found.

\section{Discussion and Conclusion}
\label{sec:Discussion & Conclusion}

In this work we explore the local stellar environment of SGR 1935+2154 using astrometry from Gaia DR3 and multi-band photometry, and find that the environment is a quiet one. In the following, we outline the key results and discuss their implications.

\begin{enumerate}
    \item 
    We search for stars within a 0.87-degree ($\sim 100\, d_{\rm \,6.6 kpc}\rm\, pc$) radius centered around SGR 1935+2154, remove foreground sources, and find no concentration of massive stars. At most 5 massive stars would be present within 6.6 kpc towards SGR 1935+2154, increasing to 9 within 10 kpc and 26 within 15 kpc, corresponding to surface densities of $2\times10^{-4}$, $1\times10^{-4}$, and $2\times10^{-4}\,\mathrm{pc}^{-2}$, respectively, which are only a quarter of the surface densities of massive stars in the solar neighborhood. We note that these numbers of massive stars towards SGR 1935+2154 should be regarded as upper limits, because they are obtained assuming that all massive star candidates we found are genuine massive stars associated with the in situ environment of SGR 1935+2154, without considering differences in extinction and distance, or the chance that some may in fact be lower-mass foreground stars. 
     
    \item 
    We backtrack the trajectories of SGR 1935+2154 and its surrounding stars to search for potential massive companions, but find no such companions. Over the time span of $10,000$ to $100,000$ yr, covering the likely age range of SGR 1935+2154, the closest massive star candidate is located about $450^{\prime \prime}$ (median) or $123^{\prime \prime}$ (minimum), corresponding to $14\,d_{\rm \,6.6 kpc}\,\rm pc$ or $4\,d_{\rm \,6.6 kpc}\,\rm pc$, from SGR 1935+2154. 
\end{enumerate}

Given the quiet surroundings of SGR 1935+2154, the progenitor is unlikely to be a very massive star formed in situ. Instead, the progenitor may have been one of the following three possibilities, listed here in decreasing order of likelihood:
\begin{enumerate}
    \item A non-very massive star formed in situ or ex situ. It is possible for stars other than very massive ones to produce magnetars \citep[e.g.,][]{2009ApJ...707..844D, 2022ApJ...926..111W}. If SGR 1935+2154 falls into this case and its progenitor formed in situ, it provides a natural explanation for the quiet local environment. The progenitor could also be an ex situ runaway or walkaway star (in or not in a binary system), but if so, our current study alone cannot provide enough information about its birthplace, and further investigation is required.

    \item A merger of a stellar system or AIC of a WD formed in situ or ex situ. Binary NS mergers \citep{2013ApJ...771L..26G}, binary WD mergers \citep{2006MNRAS.368L...1L}, NS-WD mergers \citep{2020ApJ...893....9Z}, and AIC of WDs \citep{1991ApJ...367L..19N} may form magnetars. However, we note that studies of the host galaxies of FRBs \citep[e.g.,][]{2021ApJ...907L..31B, 2023ApJ...954...80G} suggest that most of the magnetars are formed via CC supernovae, and all channels in addition to the CC may provide less than 1\% of NSs \citep{2023Parti...6..451P}. Besides the above scenarios, CC resulting from binary mergers may also generate magnetars, which can have a long delay-time \citep{2017A&A...601A..29Z}. This channel could account for a notable portion of the magnetar population \citep{2025arXiv251106554H}. In this case, the travel time from binary birth to magnetar formation may exceed 100 Myr, so the progenitor could lie outside our search region, necessitating further investigation.
    
    \item A very massive star formed ex situ. In the binary supernova scenario \citep[BSS, e.g.,][]{2019A&A...624A..66R}, if the progenitor is a $\geq 25\,\rm M_{\odot}$ companion ejected at the primary's CC, its travel time to the end of life is $<2$ Myr, even for a 40 $\rm M_{\odot}$ primary that exploded quickly \citep[single star lifetimes from][]{1998A&A...334..505P}. This implies that it must be a runaway star with a velocity exceeding $50\,d_{\rm \,6.6 kpc}\,\rm km\,s^{-1}$, given our $100\,d_{\rm \,6.6 kpc}\,\rm pc$ search radius. The runaway $(\geq 30\,\rm km\,s^{-1})$ probability among the total binaries that the companion is in the main sequence at the primary's CC is $0.78\times0.86\times0.75\times0.05\approx2.5\%$ \citep[Fig. 4 in][]{2019A&A...624A..66R}, while that for velocity exceeding $50\,\rm km\,s^{-1}$ is much lower \citep[Fig. 5 in][]{2019A&A...624A..66R}, making this scenario highly unlikely. Two other possibilities are that the progenitor was ejected through dynamical ejection from a star cluster \citep{1967BOTT....4...86P} or that a massive binary was first ejected from a star cluster, with the progenitor being the companion in the binary, and a second acceleration of the progenitor during the primary's CC \citep[``two-step-ejection'',][]{2010MNRAS.404.1564P}. \citet{2024ApJ...966..243P} use the field OB and OBe stars in the Small Magellanic Cloud to estimate the frequencies of runaway and walkaway stars caused by the BSS, the dynamical ejection scenario, and the two-step-ejection scenario, and discover that they differ by a factor of several (see Table 4 therein). If this holds for SGR~1935+2154's birth environment, both scenarios are also highly improbable.

\end{enumerate}

Although SGR~1935+2154 is likely associated with SNR~G57.2+0.8 \citep{2014GCN.16533....1G}, the progenitor-mass constraint presented here is not based on the SNR itself, but on the surrounding stellar environment. Even so, our result agrees well with previous SNR-based studies \citep{2019A&A...629A..51Z, 2023ApJ...950..137N}, which consistently favor non-very massive progenitors. SGR~1935+2154 therefore represents an important independent case: despite being associated with an SNR, its progenitor is constrained here through a different method, and this independent approach again points to a non-very massive progenitor. This adds further support to the view that such progenitors may account for a considerable fraction of the magnetar population.

Finally, we note that our study focuses on hot, massive stars. Cooler late-stage massive stars are not considered, yet our results are robust as massive stars spend the vast majority of their lifetimes at high temperatures.

\begin{acknowledgments}
We thank Yi-Xuan Shao for valuable discussions on the properties of SGR 1935+2154. We also thank Bing Yan for helpful discussion about SED fitting. W.L.H.\ and P.Z.\ acknowledge the support from National Natural Science Foundation of China (NSFC) grant No.\ 12273010, the China Manned Space Program with grant Nos.\ CMS-CSST-2025-A14 and CMS-CSST-2025-A18, and the Fundamental Research Funds for the Central Universities with grant No.\ KG202502. B.Q.C.\ acknowledges the support from NSFC grant Nos.\ 12173034 and 12322304. 

This work has made use of data from the European Space Agency (ESA) mission Gaia (\url{https://www.cosmos.esa.int/gaia}), processed by the Gaia Data Processing and Analysis Consortium (DPAC, \url{https://www.cosmos.esa.int/web/gaia/dpac/consortium}); the Pan-STARRS1 Surveys, made possible by contributions from participating institutions and funding agencies; the WISE and NEOWISE missions, funded by the National Aeronautics and Space Administration; the Two Micron All Sky Survey, funded by the National Aeronautics and Space Administration and the National Science Foundation; the UKIRT Infrared Deep Sky Survey; and the IGAPS survey, which combines the IPHAS and UVEX surveys conducted with the Isaac Newton Telescope. This research also made use of the SVO Filter Profile Service ``Carlos Rodrigo''.

We thank the ChatGPT (\url{https://chatgpt.com/}) and DeepSeek (\url{https://www.deepseek.com}) language models for their help in improving the clarity and readability of this manuscript. Every incorporation of the suggested polish was cross-checked and adjusted by the author before implementation. All scientific reasoning and results were produced independently by the authors.
\end{acknowledgments}

\software{NumPy \citep{2020Natur.585..357H},  
          SciPy \citep{2020NatMe..17..261V},
          Matplotlib \citep{2007CSE.....9...90H}
          }

\appendix

\section{Retention Criteria for $BP$ and $RP$-band Data}
\label{sec: appendix A} 
We set the retention criteria for $BP$ and $RP$-band mainly based on \citet{2021A&A...649A...3R}. We refer to their $C^{*}$ index and the reliable threshold of the magnitude of $BP$. To preserve reliable $BP$ and $RP$-band data (i.e., mean flux divided by its error is large) as much as possible, we also consider the situations beyond the analyzes in \citet{2021A&A...649A...3R}, i.e., sources those are not ``gold sources'' with $-1 \leq BP - RP \leq 7$.

The retention criteria for $BP$ are defined as follows. For ``gold sources'' (phot\_proc\_mode $=$ 0) with $-1 \leq BP - RP \leq 7$, $BP$ data is retained when $BP < 20.3$ and $C^{*} \leq 3\sigma_{C^{*}}$. For ``silver or bronze sources'' (phot\_proc\_mode $\ne$ 0), as well as ``gold sources'' with $BP - RP < -1$ or $BP - RP > 7$, $BP$ data is retained when $BP < 20.3$ and phot\_bp\_mean\_flux\_over\_error $>$ 5. 

The retention criteria for $RP$ are defined as follows. For ``gold sources'' with $-1 \leq BP - RP \leq 7$, $RP$ data is retained when $BP < 20.3$ and $C^{*} \leq 3\sigma_{C^{*}}$; if $BP > 20.3$, the requirements are phot\_rp\_mean\_flux / phot\_bp\_mean\_flux $>$ 5 and $C^{*} \leq 3\sigma_{C^{*}}$. For ``silver or bronze sources'', as well as ``gold sources'' with $BP - RP < -1$ or $BP - RP > 7$, $RP$ is retained when phot\_rp\_mean\_flux\_over\_error $>$ 5.

\bibliography{sample631}{}
\bibliographystyle{aasjournal}

\end{document}